\documentclass[11pt,a4paper]{article}
\pdfoutput=1
\usepackage[]{inputenc,graphicx,amsmath,textcomp,amssymb,appendix}
\usepackage{subfigure}  % use for side-by-side figures
\newcommand{\be}{\begin{equation}}
\newcommand{\ee}{\end{equation}}
\newcommand{\bea}{\begin{eqnarray}}
\newcommand{\eea}{\end{eqnarray}}

\catcode`@=12

\begin{document}
\begin{flushright}
%SINP-APC-13/02
\end{flushright}
\thispagestyle{empty}
\begin{center}
{\Large\bf 
{ Low Energy Gamma Ray Excess Confronting a Singlet Scalar Extended 
Inert Doublet Dark Matter Model }}\\
\vspace{1cm}
{{\bf Amit Dutta Banik} \footnote{email: amit.duttabanik@saha.ac.in}, 
{\bf Debasish Majumdar} \footnote{email: debasish.majumdar@saha.ac.in}}\\
\vspace{0.25cm}
{\normalsize \it Astroparticle Physics and Cosmology Division,}\\
{\normalsize \it Saha Institute of Nuclear Physics,} \\
{\normalsize \it 1/AF Bidhannagar, Kolkata 700064, India}\\
\vspace{1cm}
%%%%%%%%%%%%%%%%%%%%%%%%%%%%%%%%%%%%%%%%%%%%%%%%%%
%{\bf ABSTRACT}
%%%%%%%%%%%%%%%%%%%%%%%%%%%%%%%%%%%%%%%%%%%%%%%%%%
\end{center}
\begin{abstract}
Recent study of gamma rays originating from the region of galactic centre has
confirmed an anomalous $\gamma$-ray excess within the energy range 1-3 GeV.
This can be explained as the consequence of pair annihilation of a 31-40 GeV
dark matter into $b \bar b$ with thermal annihilation cross-section
$\sigma v \sim 1.4-2.0 \times 10^{-26}~\rm{cm^3/s}$. In this work we revisit
the Inert Doublet Model (IDM) in order to explain this gamma ray excess. Taking
the lightest inert particle (LIP) 
as a stable DM candidate we show that a 31-40 GeV dark matter derived from IDM
will fail to satisfy experimental limits on dark matter direct detection 
cross-section obtained from ongoing direct detection experiments and is also
inconsistent with LHC findings. We show that a singlet extended inert doublet 
model can easily explain the reported $\gamma$-ray excess which is as 
well in agreement with Higgs search results at LHC and other observed results 
like DM relic density and direct detection constraints.
\end{abstract}
\newpage
\section{Introduction}
Recent results from Femi-Lat data have confirmed the existence of GeV scale
$\gamma$-ray excess which appear to be emerging from the region of galactic
centre (GC) \cite{Goodenough:2009gk}-\cite{Daylan:2014rsa}. 
The annihilation of dark matter at the galactic centre may well be a 
cause for such excesses. The $\gamma$-ray peak in the energy range 1-3 GeV of 
gamma rays observed by Fermi-Lat to have come from the direction of 
galactic centre is addressed in a recent work by Dan 
Hooper et al \cite{Daylan:2014rsa}. In that work they show that a dark matter 
candidate within the mass range of 31-40 GeV primarily 
annihilating into $b \bar b$ or a 7-10 GeV dark matter primarily 
annihilating into 
$\tau \bar \tau$ \cite{Daylan:2014rsa}-\cite{Kong:1404} that eventually 
produce gamma, can well explain this 
observed phenomenon of excess gamma in 1-3 GeV energy range.
Some works \cite{Hooper:2013rwa}-\cite{Huang:2013pda} even suggest a DM 
candidate with 
mass $61.8^{+6.9}_{-4.9}$ can also explain this observed excess when  
their annihilation cross-section  
$\langle \sigma v\rangle _{b \bar b}$ to 
$b \bar b$ is  
$\sim 3.30^{+0.69}_{-0.49}\times 10^{-26} {\rm cm^3/s}$.
%These priliminary annihilation products ultimately produce $\gamma$-rays. 
Different particle physics models are studied and proposed in the literature
in order to explain the anomalous excess of gamma ray in the energy range
$\sim$ 1-3 GeV  \cite{Kaye:1310}-\cite{Basak:1405}. 
In this work we attempt to explore whether a dark matter candidate 
within the framework
of the inert doublet model (IDM) \cite{ma}-\cite{borah} can explain
this gamma ray excess in the gamma energy region of 1-3 GeV.
In the inert doublet model, an additional scalar SU(2) doublet 
is added to the Standard Model (SM) which is assumed to develop 
no vacuum expectation value (VEV). 
An unbroken $Z_2$ symmetry ensures that the added scalar is stable  
and does not interact with the SM fermions (inert). 
The lightest stable inert particle (LIP) in this model can be 
a viable DM candidate.
%Detailed study of IDM has been extensively done in various literature.
%Apart from providing a stable DM candidate, IDM also deals with several 
%interesting phenomena like leptogenesis, neutrino mass generation and Higgs
%phenomenology.
% In this work we have explored IDM in order to explain the GeV
%scale $\gamma$-ray excess observed in the proximity of GC.
We show in this work that although LIP dark matter in IDM model 
may indeed provide a 31-40 GeV dark matter which satisfies observed DM relic
density, but this candidate (of mass $\sim 31-40$ GeV) does not 
withstand the latest bounds from dark matter direct detection 
experiments as well as the LHC bound on $R_{\gamma\gamma}$. 
We then propose in this work, an extension of this IDM model whereby 
an additional singlet scalar is added to the IDM model mentioned above.
This newly added scalar singlet acquires a non zero
VEV and mixes up with the SM Higgs, thus provides an extra scalar boson and 
scalar resonance. The LIP dark matter candidate
In this resulting extended IDM, as we show in this work, one can 
obtain an LIP dark matter candidate  
in the mass range of 31-40 GeV which sumultaneously satisfy the 
relic density bound from Planck experiment, direct detection experimental 
results and the bound on $R_{\gamma\gamma}$ from LHC experiment.
We show that the calculation of gamma ray flux obtained from the annihilation   
of such a dark matter from the extended IDM model proposed in this work 
can explain the 1-3 GeV $\gamma$-ray excess observed
by Fermi-LAT from GC region. 
The paper is organised as follows : In
Section~\ref{S:IDM}, we revisit the Inert Doublet Model of dark matter and show
that for a 31-40 GeV DM, IDM cannot satisfy the constraints obtained from recent
direct detection bounds on DM nucleon scattering cross-section $\sigma_{\rm SI}$
and is also inconsistent with the LHC constraints. In Section~\ref{S:SIDM}, we
propose the singlet extended IDM and study the viability of the model to provide
a DM candidate in the mass range 31-40 GeV that yields the right annihilation
cross-section to $b \bar b$ final state ($\langle \sigma v\rangle _{b \bar b}$)
required to explain the observed $\gamma$-ray excess in the energy range 1-3 GeV.   
%%%%%%%%%%%%%%%%%%%%%%%%%%%%%%%%%%%%%%%%%%%%%%%%%%%%%%%**********
%%investigate the singlet extended IDM and study the vaibility of the model
%%in order to provide a 31-40 GeV DM with exact 
%%$\langle \sigma v\rangle _{b \bar b}$ 
%%required to explain the excess observed in $\gamma$-ray.
%%%%%%%%%%%%%%%%%%%%%%%%%%%%%%%%%%%%%%%%%%%%%%%%%%%%%%%%% 
We constrain the model
parameter space by various experimental results such as DM relic density 
obtained from Planck, DM-nucleon scattering cross-section bound from 
XENON, LUX experiments
and bound on the SM-like scalar given by LHC.
In Section \ref{S:flux}, the gamma ray flux is calculated for the dark
matter candidate in our proposed model and is compared with the observed
results by Fermi-LAT.
Finally we summarise
the work in Scetion~\ref{S:summary}. 
\section{Dark Matter in Inert Doublet Model and Fermi-LAT observed 
gamma ray excess}
\label{S:IDM}
IDM is a simple extension of SM of particle physics which includes an 
additional Higgs doublet that acquires no VEV. The added doublet do not 
interact with the SM sector due to imposition of a discrete $Z_2$ symmetry
under which all the SM particles are even but the doublet is odd. The most
general CP conserving potential for IDM is given as,
\bea
V &=& m_{11}^2 {\Phi_H}^\dagger {\Phi_H} + m_{22}^2  {\Phi_I}^\dagger {\Phi_I} 
+ \lambda_1 ({\Phi_H}^\dagger {\Phi_I})^2
+ \lambda_2 ({\Phi_I}^\dagger {\Phi_I})^2 + \lambda_3
(\Phi_H^\dagger \Phi_H)(\Phi_I^\dagger \Phi_I) \nonumber \\ &+&
\lambda_4 (\Phi_I^\dagger \Phi_H)(\Phi_H^\dagger \Phi_I) + {1 \over 2} \lambda_5
[(\Phi_I^\dagger \Phi_H)^2 + (\Phi_H^\dagger \Phi_I)^2],
\label{1}
\eea 
where ${\Phi_H}$ is the SM Higgs doublet and ${\Phi_I}$ is the 
inert doublet assuming all the
couplings ($\lambda_i,\,\, i=1,5$) in Eq.~\ref{1} are real. After 
spontaneous symmetry breaking (SSB), $\Phi_H$ 
generates a VEV $v=246$ GeV whereas the inert doublet does not produce 
any VEV. The $Z_2$ symmetry remains unbroken. The doublets are given as
\bea
\Phi_H = \left( \begin{array}{c}
                           \chi^+  \\
        \frac{1}{\sqrt{2}}(v+h+i\chi^0)  
                 \end{array}  \right) \, ,                     
&& \Phi_I =\left( \begin{array}{c}
                           H^+   \\
        \frac{1}{\sqrt{2}}(H_0+iA_0)  
                 \end{array}  \right) \, ,
\label{2}
\eea 
where $\chi^+$ and $\chi^0$ are absorbed in $W^\pm$, $Z$ after 
spontaneous symmetry breaking. 
After SSB, the masses of various scalar particles obtained  are given as,
\bea
m_h^2&=&2 \lambda_1 v^2 \nonumber \\
m_{H^{\pm}}^{2}&=&m_{22}^{2}+\lambda_{3}\frac{v^{2}}{2}  \nonumber \\
m_{H_0}^{2}&=&m_{22}^{2}+(\lambda_{3}+\lambda_{4}+\lambda_{5})\frac{v^{2}}{2} \nonumber \\
m_{A_0}^{2}&=&m_{22}^{2}+(\lambda_{3}+\lambda_{4}-\lambda_{5})\frac{v^{2}}{2}\,\, . 
\label{3}   
\eea 
where $m_h=125$ GeV, is the mass of newly found SM Higgs boson $h$, as observed
by LHC experiments CMS \cite{cms} and ATLAS \cite{atlas}.
With $\lambda_5 < 0$, the lightest 
inert particle (LIP) $H_0$ is the stable DM candidate in the model. The potential
described in Eq.~\ref{1} must be bounded from below and the corresponding 
vacuum stability conditions are given as,
\bea
\lambda_1,\,\lambda_2 > 0\, , ~~~~~~ 
\lambda_3 + 2\sqrt{\lambda_1\lambda_2}  >  0\, ,  ~~~~~~~
\lambda_3 +\lambda_4 -|\lambda_5| + 2\sqrt{\lambda_1\lambda_2}  >  0\, .
\label{4}
\eea  
Apart from the bounds obtained from vacuum stability, there are several other
constraints on the model such as perturbative bounds requiring all the 
couplings $\Lambda_i$ to be less than 4$\pi$. From LEP \cite{lep}
experiment constraints of the
$Z$ boson decay width and charged scalar mass $m_{H^{\pm}}$, we have 
\bea
m_{H_0} + m_{A_0}  >  m_Z \,\, , \nonumber \\
m_{H^{\pm}}  >  79.3 ~\rm{GeV}.
\label{5}
\eea 
%Presence of inert charged scalar $H^{\pm}$ enriches the Higgs phenomenology
%which has been studied in literature *. 
Apart from the constraints presented
in Eqs.~\ref{3}-\ref{4}, the present DM candidate $H_0$ must 
also satisfy the correct relic abundance of DM obtained from PLANCK \cite{planck} 
\be
\Omega_{\rm DM} h^2 = 0.1199{\pm 0.0027}\,\, ,
\label{6}
\ee 
where h is the Hubble parameter in the unit of 100 km~s$^{-1}$~Mpc$^{-1}$.
Dark matter relic density is obtained by solving the Boltzmann equation for
the DM species and is given as
\bea
\frac{{\rm d} n_{H_0}}{{\rm d} t} + 3 {\rm H}n_{H_0} &=& - \langle \sigma {\rm v} \rangle (n_{H_0}^{2}-n_{H_0\rm{eq}}^{2})\,\, .
\label{7}
\eea
In Eq.~\ref{7} $\langle \sigma {\rm v} \rangle$ is the total annihilation cross-section 
of the DM summing over all possible  annihilation channels, $n_{H_0}$ is the 
number density of dark matter particle $H_0$ and $n_{H_0\rm{eq}}$ is the equilibrium
number density of the same. The Hubble parameter is denoted as H in Eq.~\ref{7}. 
For the case of low mass dark 
matter scenario ($m_{H_0}\leq m_W$, $m_W$ is the mass of $W$ boson), 
%which is of our prime interest to produce the
%$\gamma$-ray excess of exact amount, 
total annihilation cross-section of DM 
candidate $H_0$ to SM particles expressed as
\bea
\langle{\sigma {\rm{v}}}_{H_0 H_0\rightarrow f\bar f}\rangle &=&  n_c\sum_f\frac{{m^2_f}}{\pi}
\beta_f^{3}
\frac{(\lambda_L/2)^2}{(4{m^2_{H_0}}-{m^2_h})^2+\Gamma_h^2 m_h^2}\,\, . 
\label{8} 
\eea  
In Eq.~\ref{8} above, $\Gamma_h$ is the total decay width of SM Higgs boson
(including the contribution from invisible decay channel), $m_f$ is the mass of 
the fermion species involved with $\beta_f = \sqrt{1-\frac{m_f^2}{m_{H_0}^2}}$. 
The Higgs-DM coupling denoted as $\lambda_L$ in Eq.~\ref{8} is of the form 
$\lambda_L = (\lambda_3 + \lambda_4 + \lambda_5)$ and $n_c$ is the colour 
quantum number with $n_c=3$ for quarks and $n_c=1$ for leptons respectively.
Invisible decay width of Higgs boson to DM particle as also the 
branching fraction ${\rm Br}_{\rm inv}$ for such invisible decay 
is written as 
\bea
\Gamma^{\rm{inv}}(h \rightarrow H_0H_0)&=& 
\frac{\lambda^2_{L} v^2}{64\pi m_h}\sqrt{1-\frac{4m^2_{H_0}}{m^2_h}}\, , 
\nonumber \\
{\rm Br}_{\rm inv} &=& \frac {\Gamma^{\rm{inv}}(h \rightarrow H_0H_0)}
{\Gamma_h}\, . 
\label{9}
\eea   
DM relic density is then calculated by solving the Boltzmann 
equation expressed in Eq.~\ref{7}, is given as
\bea
\Omega_{\rm{DM}}{\rm h}^2 &=& \frac{1.07\times 10^9 x_F}
{\sqrt{g_*}M_{\rm Pl}\langle \sigma {\rm v} \rangle}\,\, ,
\label{10}
\eea   
where $x_F = m_H/T_F$ is the freeze out or decoupling temperature of the DM
species $H_0$, $M_{Pl}$ is the Planck mass 
($M_{\rm Pl}=1.22\times 10^{19}$ GeV) and
$g^*$ is the number of effective degrees of freedom. The quantity $x_F$ 
(and subsequently the freeze out temperature $T_f$) can be 
obtained from the iterative solution to the equation
%Iterative solution to the equation
\bea
x_F &=& \ln \left ( \frac{m_H}{2\pi^3}\sqrt{\frac{45M_{\rm{Pl}}^2}{2g_*x_F}}
\langle \sigma \rm{v} \rangle \right )\,\, . 
\label{11}
\eea    
The relic density of the dark matter can be obtained
using Eqs.~\ref{8}-\ref{10} (and Eq.~\ref{11}) with the constraints 
given in Eqs.~\ref{4}-\ref{6}.
%Using the constraints from Eqs.~\ref{4}-\ref{6} and using Eqs.~\ref{8}-\ref{11},
%one can solve for the Boltzmann Eq.~\ref{7} for the DM candidate.
It is to be noted that in addition to the constraints mentioned above, 
the present DM 
candidate must also satisfy the DM direct detection experimental
limits provided by the experiments like XENON \cite{xenon12}, LUX \cite{lux}.
The experiments provide the upper bound of dark matter scattering 
cross-sections for different dark matter masses.  
The spin independent 
direct dark matter-nucleon scattering cross-section for the LIP 
dark matter $H_0$ of mass $M_{H_0}$ is expressed as
\bea
\sigma_{\rm {SI}}= \frac{\lambda_L^2}{4\pi}\frac{1}{m_h^4} f^2
\frac{m_N^4}{(m_{H_0}+m_N)^2},
\label{12}
\eea   
\begin{figure}[h!]
\centering
\subfigure[]{
\includegraphics[height=7 cm, width=7 cm,angle=0]{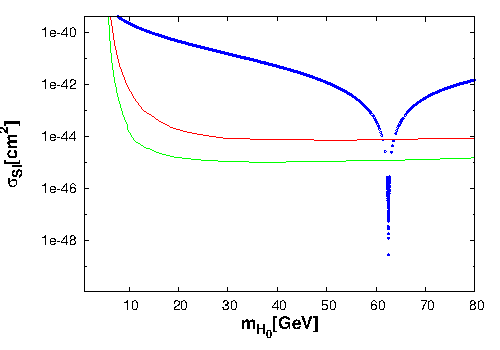}}
\subfigure []{
\includegraphics[height=7 cm, width=7 cm,angle=0]{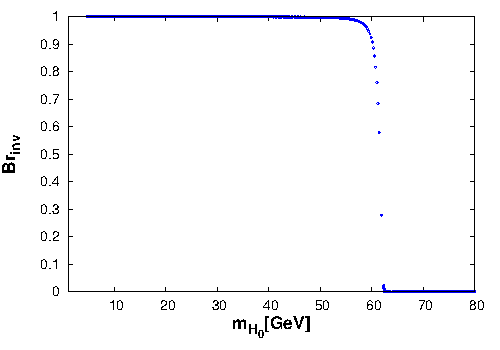}}
\caption{The left panel shows the $m_{H_0}-\sigma_{\rm SI}$ space allowed by DM 
relic density obtained from PLANCK. The right panel presents the variation of
invisible decay branching ratio ${\rm Br}_{\rm inv}$ with DM mass $m_{H_0}$ for the same.}
\label{fig1}
\end{figure}
where $m_N$ is the mass of scattering nucleon and $f$ is related to the matrix 
element of Higgs-nucleon coupling is taken to be $\simeq 0.3$ \cite{hall}.
We further 
restrict the allowed model parameter space by assuming the invisible decay 
branching ratio of SM Higgs ${\rm Br}_{\rm inv} < 20\%$ \cite{Belanger}. The 
branching ratio ${\rm Br}_{\rm{inv}}$ is the ratio 
of the Higgs
invisible decay width to the total Higgs decay width as discussed below. 
We compute, using Eq.~\ref{12} and with the constraints given in 
Eqs.~\ref{4}-\ref{6}, the LIP dark matter scattering 
cross-section, $\sigma_{\rm SI}$ for different values of LIP dark matter 
mass, $m_{H_0}$. It is therefore ensured that these calculations are 
performed for those LIP dark matter masses for which the relic 
density criterion (Eq.~\ref{6}) is satisfied.  
The results are plotted in Fig.~\ref{fig1}a 
(in $\sigma_{\rm SI}~ - ~ m_{H_0}$ plane). Superimposed 
on this plot in Fig.~\ref{fig1}a are the  
the bounds obtained from XENON100 (red line) and LUX (green line) 
experimental results for comparison. 
It is clear from Fig.~\ref{fig1}a that an LIP dark matter within the 
framework of IDM does not have a mass region in the range
31-40 GeV that satisfies the allowed bounds given by 
both the XENON100 and LUX experiments in   
$\sigma_{\rm SI}~ - ~ m_{H_0}$ plane. 
One may recall that the previous analysis to explain the 
Fermi-LAT $\gamma$-ray excess in the gamma ray energy range 
of $1-3$ GeV \cite{Daylan:2014rsa} from the aniihilation of dark matter at the 
galctic centre requires a dark matter candidate having mass 
in the range $31-40$ GeV.    
We also compute the Higgs 
invisible decay branching ratio ${\rm Br}_{\rm inv}$ for different 
$m_{H_0}$ using Eq.~\ref{9} imposing the same constraints as above 
(Eqs.~\ref{4}-\ref{6}) and the results are plotted in Fig.~\ref{fig1}b.
It is also evident from Fig.~\ref{fig1}b that the LIP mass 
($m_{H_0}$) in the range 31-40 GeV  
does not satisfy the ${\rm Br}_{\rm inv}$ limit of  
${\rm Br}_{\rm inv} < 20\%$.  
Thus from both Fig 1a and Fig 1b, it can be concluded that an LIP 
dark matter in the inert doublet model cannot account for a viable 
dark matter candidate in the mass range of 31-40 GeV.

However, from Fig 1a and 1b, it is clear we have a viable dark matter candidate
in the IDM framework in the region of Higgs resonance 
with mass ($m_{H_0}\simeq m_h/2$) that not only satisfies the relic density 
bound for dark matter but also is consistent with DM direct 
detection results and the bounds for Higgs invisible decay as well. 
Earlier model independent analysis  
\cite{Hooper:2013rwa}-\cite{Huang:2013pda} have reported that a  
dark matter with mass near Higgs resonance
can produce the observed excess of $\gamma$-ray in the gamma energy range
$1-3$ GeV if the secondary $\gamma$-ray is produced out of the 
primary annihilation process ${\rm DM}~{\rm DM} \rightarrow b \bar{b}$
with the annihilation cross-section  
$\langle \sigma v \rangle_{b \bar b} \sim 3.30^{+0.69}_{-0.49}\times 
10^{-26}~{\rm cm^3/s}$.
However for IDM with mass $m_{H_0} \sim m_h/2$, the respective 
annihilation cross-scetion of LIP dark matter $H_0$ into $b \bar b$
channel is found to be
result $\langle \sigma v \rangle_{b \bar b} 
\sim 1.7 \times 10^{-26}$ which is almost half the required 
annihilation cross-section. Hence the gamma-ray flux computed 
for this LIP dark matter (with $b \bar{b}$ to be the primary 
annihilation channel)
does not comply with the observed excess in $\gamma$-ray. 
%A considerable astrophysical boost is to be assumed. 

Thus it is apparent that a viable dark matter 
candidate (mass $\sim m_h/2$) in the IDM 
model discussed so far where only an inert SU(2) doublet is added 
to SM, fails to explain the excess
gamma ray in the energy range 1-3 GeV as observed by Fermi-LAT
in the direction of galactic centre. Hence we consider a feasible
extension of the model.    

\section{Inert Doublet Model with additional singlet scalar}
\label{S:SIDM}
We modify the IDM formalism given in Sect. 2 by adding another singlet 
scalar with the model. The resulting theory now includes an inert SU(2)
doublet as before and an additional scalar singlet added to the Standard 
Model. The newly added scalar singlet
generates a VEV and is even under the discrete $Z_2$ symmetry. 
The LIP of the inert doublet is the dark matter candidate in this 
formalism too. We demonstrate that our proposed extended IDM 
provides a viable LIP dark matter candidate in the mass range of 
$31-40$ GeV and the annihilation cross-section to $b \bar{b}$ channel  
for such a candidate can be calculated to be in the right ball park required 
to explain the excess $\gamma$ peak from GC seen by Fermi-LAT in 1-3 GeV 
energy range and is also consistent with the LHC constraint.

The most
general potential for the model is 
\bea
V &=& m_{11}^2 {\Phi_H}^\dagger {\Phi_H} + m_{22}^2  {\Phi_I}^\dagger {\Phi_I} + {1 \over 2}
m_s^2 S^2 + \lambda_1 ({\Phi_H}^\dagger {\Phi_H})^2
+ \lambda_2 ({\Phi_I}^\dagger {\Phi_I})^2 + \lambda_3
(\Phi_H^\dagger \Phi_H)(\Phi_I^\dagger \Phi_I) \nonumber \\ &+&
\lambda_4 (\Phi_I^\dagger \Phi_H)(\Phi_H^\dagger \Phi_I) + {1 \over 2} \lambda_5
[(\Phi_I^\dagger \Phi_H)^2 + (\Phi_H^\dagger \Phi_I)^2] + \rho_1
(\Phi_H^\dagger \Phi_H) S + \rho'_1 (\Phi_I^\dagger \Phi_I) S \nonumber
\\  &+&  \rho_2 S^2 (\Phi_H^\dagger \Phi_H) +  \rho'_2
S^2 (\Phi_I^\dagger \Phi_I) +{1 \over 3} \rho_3 S^3 + {1 \over 4} \rho_4 S^4 ,
\label{13}
\eea  
where $\Phi_H$ and $\Phi_I$ are the same as in Eq.~\ref{1} with
$S= s+v_s$, $v_s$ being the VEV of the singlet scalar. All the parameters in
Eq.~\ref{13} are assumed to be real. The newly added scalar singlet $s$ mixes 
with the SM Higgs $h$ resulting in two physical scalar bosons $h_1$ and $h_2$
and they are expressed as,
\bea 
h_1=h~\cos\alpha - s~\sin\alpha \, , \nonumber \\
h_2=h~\sin\alpha + s~\cos\alpha \, , 
\label{14}
\eea   
where $\alpha$ is the angle of mixing. Minimising the potential in 
Eq.~\ref{13} we obtain the conditions,
\bea
m_{11}^2+\lambda_1 v^2+ \rho_1 v_s + \rho_2 v_s^2 = 0 \, ,\nonumber \\ 
m_s^2+ \rho_3 v_s + \rho_4 v_s^2 + \frac{\rho_1 v^2}{2v_s} + \rho_2 v^2 =0 \, .
\label{15}
\eea 
The mass terms for the scalars can be obtained as 
\bea
\mu_h^2&=&2 \lambda_1 v^2 \nonumber \\
\mu_s^2&=&\rho_3 v_s + 2\rho_4 v_s^2 - \frac{\rho_1 v^2}{2 v_s} \nonumber \\
\mu_{hs}^2&=&(\rho_1+ 2 \rho_2 v_s) v  \nonumber \\
m_{H^{\pm}}^{2}&=&m_{22}^{2}+\lambda_{3}\frac{v^{2}}{2}+\rho'_1 v_s+\rho'_2 v_s^2  \nonumber \\
m_{H_0}^{2}&=&m_{22}^{2}+(\lambda_{3}+\lambda_{4}+\lambda_{5})\frac{v^{2}}{2}+\rho'_1 v_s+\rho'_2 v_s^2 \nonumber \\
m_{A_0}^{2}&=&m_{22}^{2}+(\lambda_{3}+\lambda_{4}-\lambda_{5})\frac{v^{2}}{2}
+\rho'_1 v_s+\rho'_2 v_s^2\,\, . 
\label{16}   
\eea  
As in Sect. 2, the lightest inert particle or LIP is $H_0$ 
when $\lambda_5 < 0$ and is the candidate for dark matter in this 
extended IDM formalism also.   

Masses of physical scalars $h_1$
and $h_2$ derived using the mass matrix are,
\be
m^2_{1,2}=\frac{\mu_h^2+\mu_s^2}{2}\pm\frac{\mu_h^2-\mu_s^2}{2}\sqrt{1+x^2},
\ee
\label{17}
where $x=\frac{2\mu^2_{hs}}{(\mu^2_h-\mu^2_s)}$.
We consider $h_2$ with mass $m_2$ to be the SM-like Higgs boson having
mass 125 GeV and we assume $m_2>m_1$ where $m_1$ is the mass of the 
singlet scalar. Vacuum stability conditions 
for this singlet extended IDM are given as \cite{kannike},
\bea
\lambda_1,\,\lambda_2,\,\rho_4 > 0\, , ~~~~~~ 
\lambda_3 + 2\sqrt{\lambda_1\lambda_2}  >  0\, ,  ~~~~~~~
\lambda_3 +\lambda_4 -|\lambda_5| + 2\sqrt{\lambda_1\lambda_2} & > & 0\, ,
\nonumber\\
\rho_2 +\sqrt{\lambda_1\rho_4}  >  0\, , ~~~~~~~~~~~~ 
\rho_2' +\sqrt{\lambda_2\rho_4}  >  0\, ,~~~~~~~~~~~~
\nonumber \\
2 \rho_2 \sqrt{\lambda_2} +2 \rho_2' \sqrt{\lambda_1} +  \lambda_3 \sqrt{\rho_4} \hskip12pt\nonumber \\
+ 2 \left(\sqrt{\lambda_1\lambda_2\rho_4} \right.
\left.
+ \sqrt{\left(\lambda_3 + 2\sqrt{\lambda_1\lambda_2}\right)
\left(\rho_2 +\sqrt{\lambda_1\rho_4}\right)
\left(\rho_2' +\sqrt{\lambda_2\rho_4}\right)}\right) & > & 0\,
\nonumber \\ 
2 \rho_2 \sqrt{\lambda_2} +2 \rho_2' \sqrt{\lambda_1} +  (\lambda_3+\lambda_4-\lambda_5) \sqrt{\rho_4} \hskip12pt\nonumber \\
+ 2 \left(\sqrt{\lambda_1\lambda_2\rho_4} \right.
\left.
+ \sqrt{\left(\lambda_3+\lambda_4-\lambda_5 + 2\sqrt{\lambda_1\lambda_2}\right)
\left(\rho_2 +\sqrt{\lambda_1\rho_4}\right)
\left(\rho_2' +\sqrt{\lambda_2\rho_4}\right)}\right) & > & 0\,.
\label{18}
\eea
%\bea
%\lambda_1,\,\lambda_2,\,\rho_4 > 0\, , ~~~~~~ 
%\lambda_3 + 2\sqrt{\lambda_1\lambda_2}  >  0\, ,  &&
%\lambda_3 +\lambda_4 -|\lambda_5| + 2\sqrt{\lambda_1\lambda_2}  >  0\, ,
%\nonumber\\
%\rho_2 +\sqrt{\lambda_1\rho_4}  >  0\, , ~~~~~~~~~~~~ &&
%\rho_2' +\sqrt{\lambda_2\rho_4}  >  0\, ,~~~~
%\nonumber \\
%2 \rho_2 \sqrt{\lambda_2} +2 \rho_2' \sqrt{\lambda_1} +  
%\lambda_3 \sqrt{\rho_4}+ && \nonumber \\
%&& 2 \left(\sqrt{\lambda_1\lambda_2\rho_4} \right.
%\left.
%+ \sqrt{\left(\lambda_3 + 2\sqrt{\lambda_1\lambda_2}\right)
%\left(\rho_2 +\sqrt{\lambda_1\rho_4}\right)
%\left(\rho_2' +\sqrt{\lambda_2\rho_4}\right)}\right) & > & 0\, .
%\label{18}
%\eea
Imposing the vacuum stability conditions (Eq.~\ref{18}) and applying the 
perturbative bounds and
constraints from Eqs.~\ref{5}-\ref{6} we solve the Boltzmann equation 
in Eq.~\ref{7}. Note that, for the proposed extended IDM model, 
both the annihilation cross-section 
$\langle{\sigma {\rm{v}}}_{H_0 H_0\rightarrow f\bar f}\rangle$
and the invisible decay width  
$\Gamma^{\rm{inv}}_i(h_i \rightarrow H_0H_0)$ must be modified. 
The thermal averaged annihilation
cross-section for the LIP dark matter in the present model is expressed as
\bea
\langle{\sigma {\rm{v}}}_{H_0 H_0\rightarrow f\bar f}\rangle &=&  n_c \sum_f\frac{{m^2_f}}{\pi}
\beta_f^{3}
\left|\frac{\lambda_{h_1H_0H_0}\cos{\alpha}}{4{m^2_{H_0}}-{m^2_1}+i\Gamma_1 m_1}
+\frac{\lambda_{h_2H_0H_0}\sin{\alpha}}{4{m^2_{H_0}}-{m^2_2}+i\Gamma_2 m_2}\right|^2 \,\, . 
\label{19}
\eea 
In Eq.~\ref{19} above, $\Gamma_i$ (i=1,2) is the total decay width of $h_i$
and the coupling $\lambda_{h_1H_0H_0}$, $\lambda_{h_2H_0H_0}$ are 
\bea
\lambda_{h_1H_0H_0}v=\left(\frac{\lambda_L}{2}c_{\alpha}-\frac{\lambda_s}{2}s_{\alpha}\right)v \, ,\nonumber \\
\lambda_{h_2H_0H_0}v=\left(\frac{\lambda_L}{2}s_{\alpha}+\frac{\lambda_s}{2}c_{\alpha}\right)v
\label{20} 
\eea  
with $\lambda_L=\lambda_3+\lambda_4+\lambda_5$ and $\lambda_s=\frac{\rho_1'+2 \rho_2' v_s}{v} $.
Invisible decay width of $h_1$ and $h_2$ are given as
\be
\Gamma^{\rm{inv}}_i(h_i \rightarrow H_0H_0)= \frac{\lambda^2_{h_iH_0H_0} v^2}{16\pi m_i}\sqrt{1-\frac{4m^2_{H_0}}{m^2_i}}\, . 
\label{21}
\ee
The LIP-nucleon spin independent (direct detection) cross-section  
in this singlet scalar extended IDM is modified as, 
\bea
\sigma_{\rm {SI}}= \frac{1}{\pi}\frac{m_N^4}{(m_{H_0}+m_N)^2} f^2
\left(\frac{\lambda_{h_1H_0H_0}\cos\alpha}{m_1^2}+
\frac{\lambda_{h_2H_0H_0}\sin\alpha}{m_2^2} \right)^2.
\label{22}
\eea  
As before, we restrict the model parameter space using the conditions from
vacuum stability (Eq.~\ref{18}), unitarity, LEP, DM relic density
from PLANCK. In addition,  
%and taking $Br_{\rm{inv}} < 0.2$ \cite{Belanger}.
we also take into account the
modification of signal strength of SM Higgs ($h_2$) to any particular channel
that may occur due to the mixing with other scalar ($h_1$). 
The signal strength to any specific channel is given as,
\bea
R &=& \frac {\sigma} {\sigma^{\rm SM}} \frac {\rm Br} {{\rm Br}^{\rm SM}}
\label{23}
\eea
where $\sigma$ and ${\sigma^{\rm SM}}$ are the Higgs production 
cross-section in the present model and in SM respectively whereas
Br and ${\rm Br}^{\rm SM}$ are the respective branching ratios 
to any channel for the present model and SM. 
\begin{figure}[h!]
\centering
\subfigure[]{
\includegraphics[height=7 cm, width=7 cm,angle=0]{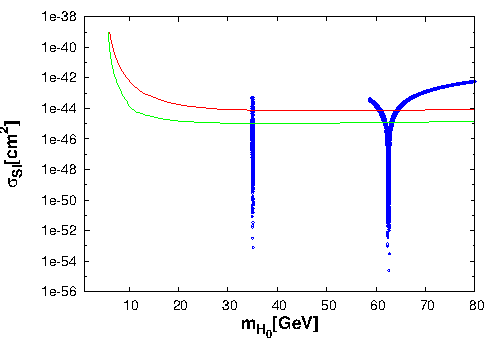}}
\subfigure []{
\includegraphics[height=7 cm, width=7 cm,angle=0]{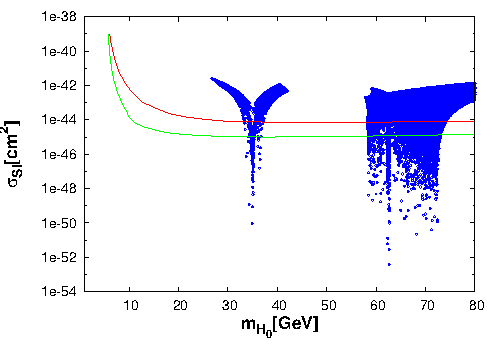}}
\subfigure[]{
\includegraphics[height=7 cm, width=7 cm,angle=0]{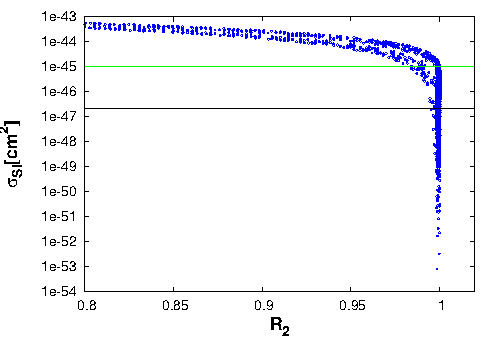}}
\subfigure []{
\includegraphics[height=7 cm, width=7 cm,angle=0]{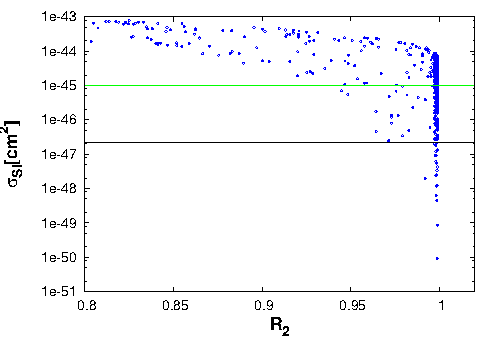}}
\caption{The upper panel shows the valid $m_{H_0}-\sigma_{\rm SI}$ plane
obtained for $m_2=70$ GeV with 
$\cos\alpha=9.0\times 10^{-3}~{\rm and}~3.5\times10^{-2}$.
The lower panel shows the variation of signal strength $R_2$ with $
\sigma_{\rm SI}$ for $m_{H_0}=35$ GeV for the same. }
\label{fig2}
\end{figure}
As the present model (extended IDM) involves two scalars $h_1$
and $h_2$, signal strengths $R_1$ and $R_2$ for both the scalars are 
given as 
\bea
R_1 = \frac{\sigma^1(pp\rightarrow h_1)}
{\sigma^{\rm SM}(pp\rightarrow h_1)}
\frac{{\rm Br}(h_1 \rightarrow xx)}{{\rm Br}^{\rm SM} (h_1\rightarrow xx)},   
\hskip 10pt
R_2 = \frac{\sigma^2(pp\rightarrow h_2)}
{\sigma^{\rm SM}(pp\rightarrow h_2)}
\frac{{\rm Br}(h_2 \rightarrow xx)}{{\rm Br}^{\rm SM} (h_2\rightarrow xx)}
\label{24}
\eea
where $xx$ is any SM final state with $\frac{\sigma^i}{\sigma^{\rm SM}} =
\cos^2\alpha~{\rm or}~\sin^2\alpha$ for $i=1,2$ respectively. 
Since $h_2$ is the SM-like
scalar with mass $m_2=125$ GeV, we take $R_2\geq 0.8$ \cite{atlas3}
for SM-like scalar to satisfy LHC results.
It is to be noted that some of the channels ($\gamma Z,~\gamma \gamma$) will
suffer considerable changes due to the presence of inert charged scalars 
($H^{\pm}$) addressed in \cite{arhrib,maria,goudelis,banik}.
Effect of the charged scalars on those channels 
are also taken into account (see Appendix A). We put further bound on model 
parameter space from the experimental limits for Higgs to diphoton signal 
strength
$R_{\gamma \gamma}$ given by ATLAS \cite{atlas1} and CMS \cite{cms1}.
Our calculation yields that for the allowed parameter space obtained from 
vacuum stability, relic density, LEP constraints as also with the condition 
$R_2\geq 0.8,~{\rm Br}_{\rm{inv}}\leq0.2$,
the Higgs to diphoton
signal strength predicted by ATLAS is not favoured by the present model
and hence we constrain
the model with the experimental value of $R_{\gamma \gamma}$ only from CMS 
experiment.
Taking all these constarints into account, 
we now compute the LIP dark matter 
(in extended IDM) scattering cross-sections 
$\sigma_{\rm SI}$ (Eq.~\ref{24}) for the LIP masses ($m_H$)  
for two different mixing angles $\alpha$ given by 
$\cos \alpha = 9.0\times 10^{-3}~{\rm and}~3.5\times10^{-2}$.
The results for two chosen mixing angles are plotted in 
Fig.~\ref{fig2}a and Fig.~\ref{fig2}b respectively in $m_{H_0}-\sigma_{\rm SI}$ 
parameter space. The calculations are performed with a chosen value 
$m_1 = 70$ GeV for the mass of the scalar singlet $h_1$. 
%The choice of $m_1$ is consistent with the observed LHC signals
%which is discussed latter in.
Diret detection bounds from XENON100 and LUX are shown in Fig.~\ref{fig2}a-b 
with the same color definitions used in Fig.~\ref{fig1}a.
\begin{figure}[h!]
\centering{
\includegraphics[height=7 cm, width=7 cm,angle=0]{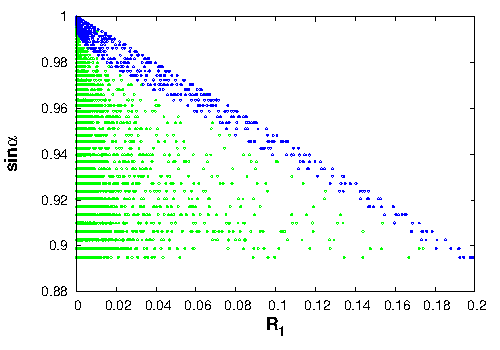}}
\caption{Allowed parameter space in $R_1-\sin\alpha$ plane for $m_2=70$ GeV.
Also shown in blue corresponds to the parameter space for $m_{H_0}=35$ GeV.}
\label{fig3}
\end{figure} 
It is clear from Fig.~\ref{fig2}a-b that
apart from obtaining a LIP dark matter of mass $\sim m_2/2$ (Higgs resonance) 
allowed by both XENON100 and LUX, we also obtain another allowed LIP mass of
35 GeV (due to the resonance of the added scalar involved
in the model). Thus, the present modified inert doublet model  
produces a viable DM candidate with a mass of 35 GeV. 
Figs.~\ref{fig2}a-b also indicate that the resonant behaviour is prominent
for smaller values of mixing angle $\alpha$. Increase in the mixing angle
broadens the allowed $m_{H_0}-\sigma_{\rm SI}$ parameter space with appreciable 
increase in DM-nucleon cross-section. 
%\begin{figure}[h!]
%\centering
%\subfigure[]{
%\includegraphics[height=7 cm, width=7 cm,angle=0]{35a.png}}
%\subfigure []{
%\includegraphics[height=7 cm, width=7 cm,angle=0]{35b.png}}
%\caption{The left panel shows the range of $\lambda_L-\lambda_{h_2h_0h_0}$ parameter
%space obtained for $m_{H_0}=35$ GeV with $m_2=70$ GeV whereas the right panel depict
%the same in $\lambda_s-\lambda_{h_1H_0H_0}$ plane. The region in blue satisfies
%LUX direct detection bound for $m_{H_0}=35$ GeV.}
%\label{fig4}
%\end{figure}
In Fig.~\ref{fig2}c-d we show the variation of $R_2$ with $\sigma_{\rm SI}$
where LIP dark matter mass $m_{H_0}=35$ GeV is considered for the  
two mixing angles as chosen for Fig.~\ref{fig2}a-b.
%for the same values of mixing angle.
Horizontal lines in green and black are the values of 
$\sigma_{\rm SI}$ as obtained from the allowed regions from 
LUX \cite{lux} and XENON1T \cite{xe1T} respectively for the dark matter mass
of 35 GeV. Fig.~\ref{fig2}c shows that 
as $R_2$ approaches
to unity  there is a sharp decrease in $\sigma_{\rm SI}$. 
%Also $R_2\simeq 1$
%will result in a decrease in invisible decay branching fraction.
A similar conclusion also follows from the nature of Fig. \ref{fig2}d. 
Observation of Fig. \ref{fig2}c-d reveals that a 35 GeV DM satisfying
relic density obtained from PLANCK  and direct detection bounds 
from LUX and XENON1T does not affect
the signal strength ($R_2 \sim 1$) of the SM Higgs observed in LHC. 
Fig.~\ref{fig2}c-d clearly demonstrate that the presence of a 
low mass scalar is 
necessary in order to achieve a DM of mass$\sim35$ GeV that (a) satisfy PLANCK
relic density result, (b) agree with the latest dark matter 
direct detection experimental 
bounds and also (c) yields the experimental bound for
Higgs invisible decay.
%with correct order of
%$\sigma_{SI}$ compatible with bounds on invisible decay.

Since the model 
involves an additional scalar of low mass, yet undetected by LHC, the 
corresponding signal strength for that singlet like scalar must remain small
compared to that of $h_2$. In order to demonstrate this, we compute 
the signal strength $R_1$ (Eq.~\ref{24}) for different values of 
the mixing angle $\alpha$. 
In Fig.~\ref{3} we plot the results in 
$R_1-\sin\alpha$ plane for low mass DM ($\leq m_W$). These results
satisfy the conditions $R_2\geq 0.8$ \cite{atlas3}
and ${\rm Br}_{\rm{inv}}\leq 0.2$ \cite{Belanger} with $m_1=70$ GeV and also 
consistent with relic density
reported by PLANCK. Scattered blue region in Fig.~\ref{fig3} corresponds to 35
GeV DM mass ($m_{H_0} = 35$ GeV) with 
$<\sigma v>_{b \bar b}~\sim(1.62-1.68)\times 10^{-26} 
{\rm cm^3/s}$. We show latter in this in Sec. \ref{S:flux} that such a value for
$<\sigma v>_{b \bar b}$ in case of a dark matter mass of 35 GeV 
can indeed explain the Fermi-LAT observed excess of $\gamma$-ray in 
the energy range of  1-3 GeV.    
Variation of $\sin\alpha$ with $R_1$ in Fig.~\ref{fig3} depicts 
that for the parameter space constrained by different experimental 
and theoretical bounds, the value of the  
signal strength $R_1$ remains small ($\leq 0.2$). Therefore,  
non-observance of such a scalar by LHC is justified and can possibly 
be probed in future experiment. 

\section{Calculation of gamma ray flux}
\label{S:flux}
\begin{figure}[h!]
\centering
{
\includegraphics[height=7 cm, width=7 cm,angle=0]{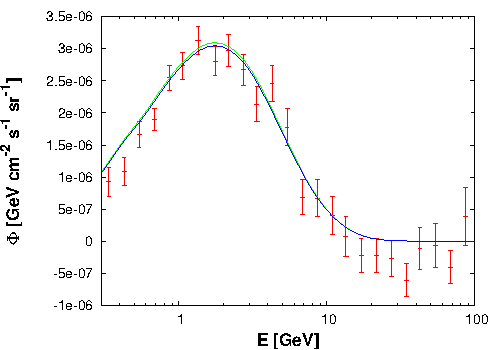}}
\caption{$\gamma$-ray flux obtained from the benchmark points in Table\ref{tab1} 
and compared with the results from \cite{Daylan:2014rsa}.}
\label{fig4}
\end{figure}

In this section we calculate the gamma ray flux from the 
galctic centre due to the annihilation 
of 35 GeV dark matter in the 
extended IDM discussed in Sect.\ref{S:SIDM}. 
The gamma ray flux produced from DM annihilation in galactic centre 
is given by 
\bea
\Phi=\frac{\langle \sigma v\rangle}{8\pi m_{DM}^2}
\frac{dN}{dE_{\gamma}} J(\psi)\,\, .
\label{25}
\eea
In Eq.~\ref{25}, $\langle \sigma v\rangle$ is the annihilation cross-section,
$m_{DM}$ is the mass of the dark matter ($m_{H_0}$ in the present scenario), 
$\frac{dN}{dE_{\gamma}}$ is the 
spectrum of photon produced due to DM annihilation.
The factor $J(\psi)$ in Eq. \ref{25} is the line of sight integral given as
\bea
J(\psi)=\int_{\rm los}\rho^2(l,\psi)dl\, ,
\label{26}
\eea
where $\psi$ is angle between the line of sight of an observer at Earth
at a distance $\ell$ from the GC and the direction from GC to Earth, 
$l$ is the distance from line of sight.
We use the generalised NFW \cite{nfw} halo profile for the DM distribution $\rho(r)$
given by
\bea
\rho(r)=\rho_0\frac{{r/r_s}^{-\gamma}}{{1+r/r_s}^{3-\gamma}}\, .
\label{27}
\eea 
In Eq.~\ref{27}, $\rho_o= 0.3~{\rm{GeV~cm^{-3}}}$ is the
local DM density at a distance 8.5 kpc from GC. For the present work
we consider $r_s= 20$ kpc and $\gamma=1.26$ \cite{Daylan:2014rsa}. 
For the calculation of gamma ray flux using Eqs.~\ref{25} - \ref{27},
we consider two values of mixing angles given by 
$\cos \alpha = 0.9\times10^{-3}~{\rm{and}}~2.5\times10^{-2}$ 
for $m_{DM} = m_{H_0} = 35$ GeV. A chosen set of values for other 
parameters and the corresponding calculated values of 
$\langle \sigma v\rangle_{b \bar b}$ and $\sigma_{\rm{SI}}$ 
for each of these two mixing angles 
are tabulated in 
Table 1. The gamma ray flux is now calculated 
\begin{table}  
\begin{center}
\vskip 0.5 cm
\begin{tabular}{|c|c|c|c|c|c|c|c|c|c|c|}
\hline
          &          &                    &                     &              &             &                                        &                         \\
  $m_1$   &  $m_{H_0}$   &   $m_H^{\pm}$  &  $\cos\alpha$       & $\lambda_L$  & $\lambda_s$ &$\langle \sigma v\rangle_{b \bar b}$    & $\sigma_{\rm{SI}}$       \\
  in GeV  &  in GeV  &    in GeV          &                     &              &             &in $\rm{cm^3/s}$                        &    in $\rm{cm^2}$       \\ 
\hline
          &          &     174.0          &  0.9$\times10^{-3}$ &  -7.89e-05   & -7.91e-02   &1.66$\times10^{-26}$           &  4.58$\times10^{-49}$     \\
   70.0   &   35.0   &                    &                     &              &             &                               &                            \\
          &          &     110.0          &  2.5$\times10^{-3}$ &   7.87e-04   &  1.26e-02   &1.65$\times10^{-26}$           &  2.52$\times10^{-48}$     \\
\hline
\end{tabular}
\end{center}
\caption{Bencmark points of singlet extended IDM with DM mass $m_{H_0} = 35$ GeV.}
\label{tab1}
\end{table}
for the LIP dark matter in our model, in case of each of these 
two set of parameter values given in Table 1 and the 
results are plotted in Fig.~\ref{fig4}.  
In Fig.~\ref{fig4} the green and blue lines correspond to the 
mixing angles given by
$\cos \alpha = 0.9\times10^{-3}~{\rm{and}}~\cos \alpha = 2.5\times10^{-2}$ 
respectively. Also shown in Fig.~\ref{fig4}, the data points for the 
observed $\gamma$-ray by Fermi-LAT for comparison. These data points 
are obtained from Ref. \cite{Daylan:2014rsa}. Fig.~\ref{fig4} clearly 
demonstrates that the viable LIP  
DM candidate in our model can very well explain the observed $\gamma$-ray flux
and its excess in the 1-3 GeV energy range while remain consistent 
with the bounds from LHC and DM direct search 
experiments.           
%%%%%%%%%%%%%%%%%%%%%%%%%%%%%%%%%%%%%%%%%%%%%%%%%%%%%%%%%%%%%%%%%%%%%
\section{Summary}
\label{S:summary}
In this paper we have revisited the inert doublet model (IDM) of dark matter
and test the viability of the model to provide a suitable explanation for the
observed excess in low energy (1-3 GeV) $\gamma$-ray emission from GC
assumed to have originated out of the annihilation of dark matter in the mass
range 31-40 GeV DM, into $b \bar b$. We show that a dark matter candidate 
within mass range 31-40 GeV
in IDM cannot satisfy the latest direct detection bounds on DM-nucleon 
cross-section predicted by experiments like LUX or XENON100 and also is inconsistent
with the limits on Higgs invisible decay. Our calculation also yield that
although IDM can provide a DM of mass $\sim m_h/2$ ($m_h$ is the mass of SM Higgs)
that is consistent with direct detection and invisible decay bounds but eventually
fails to produce the exact value of $\langle \sigma v\rangle_{b \bar b}$
required to explain the excess emisson of $\gamma$-ray.
In order to comply with the observed $\gamma$ emission results as obtained from 
Fermi-LAT in 1-3 GeV energy range,
we extend the IDM with an additional singlet scalar and explore the
viability of the model. The extension of IDM provides an additional scalar singlet 
that mixes with the SM-Higgs. We found that prescence of a low mass singlet like
scalar in the model can yield a 31-40 GeV DM that satisfy relic density bounds 
from PLANCK and
direct detection cross-section constarints from LUX or XENON experiments that 
and also yields the right DM annihilation cross-section 
$\langle \sigma v\rangle_{b \bar b}$ that would explain
the observed excess in $\gamma$-ray. The weakly coupled singlet like scalar due to 
small mixing with SM-Higgs acquires a very small signal strength which is beyond
the present LHC detection limit and can be probed 
in future collider experiments.  
\vskip 2mm
\noindent {\bf Appendix A}
\vskip 2mm 
The inert chraged scalar will contribute to Higgs decay channels like $\gamma 
\gamma$ and $\gamma Z$ through the charged scalar loop involved in the process.
Decay widths of $h_i\rightarrow\gamma \gamma,~\gamma Z$ ($i=1,2$) are given as
\bea
\Gamma(h_i\rightarrow \gamma\gamma)&=&\frac{G_F\alpha_s^2m_i^3}{128\sqrt{2}\pi^3}
\left |c_i\left(\frac{4}{3}  F_{1/2}\left(\frac{4m_t^2}{m_i^2}\right)
+ F_1 \left(\frac{4m_W^2}{m_i^2} \right)\right)
+\frac{\lambda_{h_iH^+H^-}v^2}{2m_{H^{\pm}}^2}
F_0 \left(\frac{4m_{H^{\pm}}^2}{m_i^2}\right)
\right |^2, \nonumber \\
\Gamma(h_i\rightarrow \gamma Z) &= &\frac{G_F^2\alpha_s}{64\pi^4} m_W^2 m_i^3 \left(1-\frac{m_Z^2}{m_i^2}\right)^3
\left|-2c_i \frac{1-\frac{8}{3}s^2_W}{c_W}
F_{1/2}'\left(\frac{4m_t^2}{m_i^2},\frac{4m_t^2}{m_Z^2}\right) \right. \nonumber \\
&& \left .
-c_i F_1'\left(\frac{4m_W^2}{m_i^2},\frac{4m_W^2}{m_Z^2}\right)
+\frac{\lambda_{h_iH^+H^-}v^2}{2m_{H^{\pm}}^2}
\frac{(1-2s^2_W)}{c_W}I_1\left(\frac{4m_{H^{\pm}}^2}{m_i^2},\frac{4m_{H^{\pm}}^2}{m_Z^2}\right)
\right|^2, \nonumber
\eea
where $G_F$ is the Fermi constant and $s_W$ ($c_W$) is  
$\sin\theta_W$ ($\cos\theta_W$) with $\theta_W$ represnting the weak mixing 
angle. Factor $c_i$ in the above is given as $\cos\alpha$ or $\sin\alpha$
for $i=1,2$.
Couplings $\lambda_{h_1H^+H^-}$ and $\lambda_{h_2H^+H^-}$ in the expressions of
decay widths are of the form 
\bea
\lambda_{h_1H^+H^-}v=\left(\lambda_{3}c_{\alpha}-{\lambda_s}s_{\alpha}\right)v \, , \nonumber \\
\lambda_{h_2H^+H^-}v=\left(\lambda_{3}s_{\alpha}+{\lambda_s}c_{\alpha}\right)v. \nonumber
\eea
Various loop factors corresponding to the $h_i\rightarrow \gamma 
\gamma$ process are expressed as \cite{higgshunter,djouadi1,djouadi2},
\bea
F_{1/2}(\tau)&=&2\tau[1+(1-\tau)f(\tau)],\nonumber\\
F_1(\tau)&=&-[2+3\tau+3\tau(2-\tau)f(\tau)],\nonumber\\
F_0(\tau)&=&-\tau[1-\tau f(\tau)],\nonumber
\eea
where the function $f(x)$ is given as
\bea
f(x)=\left\{ \begin{array}{ll}
\arcsin^2\left(\frac{1}{ \sqrt{x} }\right) & {\rm{for}}~~~~x \geq 1,\\
-\frac{1}{4}\left[\log\left(\frac{1+\sqrt{1-x}}{1-\sqrt{1-x}}\right)-i\pi\right]^2  & {\rm{for}}~~~~ x<1.
\end{array} \right.
\nonumber
\eea
Similarly the loop factor for $h_i\rightarrow \gamma Z$ channel are \cite{higgshunter,djouadi1,djouadi2}
\bea
F_{1/2}'(\tau,\lambda)&=&I_1(\tau,\lambda)-I_2(\tau,\lambda),\nonumber\\
F_1'(\tau,\lambda)&=&c_W\left\{4\left(3-\frac{s^2_W}{c^2_W}\right)I_2(\tau,\lambda)+
\left[\left(1+\frac{2}{\tau}\right)\frac{s^2_W}{c^2_W}-\left(5+\frac{2}{\tau}\right)\right]I_1(\tau,\lambda)\right\}.\nonumber
\eea
Expressions of the factors $I_1$ and $I_2$ are of the form
\bea
I_1(a,b)&=&\frac{ab}{2(a-b)}+\frac{a^2b^2}{2(a-b)^2}\left[f(a)-f(b)\right]+\frac{a^2b}{(a-b)^2}\left[g(a)-g(b)\right],\nonumber\\
I_2(a,b)&=&-\frac{ab}{2(a-b)}\left[f(a)-f(b)\right].\nonumber
\eea
where $f(x)$ is same as used in $h_i\rightarrow \gamma \gamma$ channel and $g(x)$ is given as
\bea
g(x)=\left\{ \begin{array}{ll}
\sqrt{x-1}\arcsin\sqrt{\frac{1}{x}}& {\rm{for}}~~~~x \geq 1,\nonumber\\
\frac{\sqrt{1-x}}{2}\left(\log\frac{1+\sqrt{1-x}}{1-\sqrt{1-x}}-i\pi\right)&{\rm{for}}~~~~x<1.
\end{array} \right.
\eea

\end{document}